\documentclass[aps,prb,reprint,superscriptaddress,longbibliography]{revtex4-2}

\usepackage{graphicx}
\usepackage{dcolumn}
\usepackage{bm}
\usepackage{amsmath}
\usepackage{amssymb}
\usepackage{mathtools}

\usepackage[utf8]{inputenc}
\usepackage{etoolbox}
\usepackage[colorlinks=true,citecolor=blue,urlcolor=blue]{hyperref}
\usepackage{tcolorbox}
\usepackage{siunitx}
\usepackage{braket}
\usepackage{todonotes}
\usepackage{array}
\usepackage{multirow}
\usepackage{float}
\usepackage{pifont}
\usepackage[version=4]{mhchem}
\usepackage{soul}

\usepackage{newtxtext,newtxmath}

\newcommand{\iu}{\mathrm{i}}

\newcommand{\bismsel}{\ce{Bi2Se3}}
\newcommand{\bismselsp}{\ce{Bi2Se3} }
\renewcommand{\vec}[1]{\boldsymbol{#1}}
\allowdisplaybreaks

\definecolor{darkpastelgreen}{rgb}{0.01, 0.75, 0.24}
\newcommand{\cmark}{\textcolor{darkpastelgreen}{\ding{51}}}
\newcommand{\xmark}{\textcolor{red}{\ding{55}}}

\begin{document}


\title[Topology of \bismselsp nanosheets]{Topology of \bismselsp nanosheets}
\author{L. Maisel Licerán}
\email{l.maiselliceran@uu.nl}
\affiliation{Institute for Theoretical Physics and Center for Extreme Matter and Emergent Phenomena, Utrecht University, Princetonplein 5, 3584 CC Utrecht, The Netherlands}
\author{S. J. H. Koerhuis}
\affiliation{Institute for Theoretical Physics and Center for Extreme Matter and Emergent Phenomena, Utrecht University, Princetonplein 5, 3584 CC Utrecht, The Netherlands}
\author{D. Vanmaekelbergh}
\affiliation{Debye Institute for Nanomaterials Science, Leonard S. Ornsteinlaboratorium, Utrecht University, Princetonplein 1, 3584 CC Utrecht, The Netherlands}
\author{H. T. C. Stoof}
\affiliation{Institute for Theoretical Physics and Center for Extreme Matter and Emergent Phenomena, Utrecht University, Princetonplein 5, 3584 CC Utrecht, The Netherlands}

\date{\today}

\begin{abstract}
    Recently, the quantum spin-Hall edge channels of two-dimensional colloidal nanocrystals of the topological insulator \bismselsp were observed directly.
    Motivated by this development, we reconsider the four-band effective model which has been traditionally employed in the past to describe thin nanosheets of this material.
    Derived from a three-dimensional $\vec{k} \vec{\cdot} \vec{p}$ model, it physically describes the top and bottom electronic surface states at the $\Gamma$ point that become gapped due to the material's small thickness.
    However, we find that the four-band model for the surface states alone, as derived directly from the three-dimensional theory, is inadequate for the description of thin films of a few quintuple layers and even yields an incorrect topological invariant within a significant range of thicknesses.
    To address this limitation we propose an eight-band model which, in addition to the surface states, also incorporates the set of bulk states closest to the Fermi level.
    We find that the eight-band model not only captures most of the experimental observations, but also agrees with previous first-principles calculations of the $\mathbb{Z}_{2}$ invariant in thin films of varying thickness.
    The band inversion around the $\Gamma$ point, which endows the surface-like bands with topology, is shown to be enabled by the presence of the additional bulk-like states without requiring any reparametrization of the resulting effective Hamiltonian.
\end{abstract}

\maketitle

\section{Introduction}

Topological insulators (TIs), and more generally topological materials, have experienced a massive surge in interest in the last decade due to their excellent prospects for energy-efficient electronics, spintronics, and transport applications\ \cite{fu2007atopological,hasan2010colloquium,moore2010birth,qi2011topological,ryu2010topological,kong2011opportunities,hasan2011three,culcer2012transport,pesin2012spintronics,bernevig2013topological,ando2013topological,ortmann2015topological,chiu2016classification,kou2017two,xu2017topological}.
A common feature shared by most conventional TIs is the appearance of protected states at the boundaries of a finite sample, which are typically perfectly conducting and whose dispersions are linear and cross the semiconducting band gap.
In three dimensions (3D) these take the form of surface states, while in two dimensions (2D) they are realized as edge states.
The properties of these boundary modes strongly depend on which symmetries are present in a given system.
Arguably, the best known examples are the integer quantum Hall state\ \cite{klitzing1980new,haldane1988model,stone1992quantum,yoshioka2002quantum,novoselov2007room,cage2012quantum}, where chiral electrons flow without dissipation along the edge due to breaking of the time-reversal symmetry (TRS), and the quantum spin-Hall (QSH) state\ \cite{kane2005quantum,bernevig2006quantum,bernevig2006quantum2,konig2007quantum,konig2008quantum,qi2010quantum,maciejko2011quantum,qian2014quantum}, which preserves the TRS and can be seen as two spin-reversed copies of the quantum Hall state with opposite spins flowing in opposite directions.

Within the vast landscape of topological materials, bismuth selenide (\bismsel) is often quoted as a prototypical three-dimensional TI\ \cite{zhang2009topological,xia2009observation,hsieh2009tunable,zhang2010first,manjon2013high,mazumder2021brief}.
Its atomic structure comes in the form of quintuple layers (QLs) stacked on top of one another and bound together by van der Waals forces\ \cite{zhang2009topological,liu2010model,zhang2011raman,cao2013structural}.
The topological nature of \bismselsp has been extensively investigated both theoretically and experimentally, and can be traced back to the existence of a large inverted topological gap arising from the spin-orbit interaction.
This leads to topologically protected surface states at the top and bottom surfaces of planar slabs.
These have been fully characterized theoretically\ \cite{liu2010oscillatory,lu2010massive,shan2010effective,linder2009anomalous} by means of a continuum model for three-dimensional \bismselsp around the $\Gamma$ point derived from $\vec{k} \vec{\cdot} \vec{p}$ theory\ \cite{zhang2009topological,liu2010model}.
A particularly interesting topic in this regard is the transition from 3D to 2D samples of \bismsel, as the dispersion of the surface states becomes gapped due to the hybridization between the modes at both surfaces, and the corresponding gap may or may not be of topological nature.
The gapping of the top and bottom surface states has been observed experimentally via angle-resolved photoemission spectroscopy\ \cite{zhang2010crossover,sakamoto2010spectroscopic,neupane2014observation}, and addressed theoretically in some of the aforementioned studies\ \cite{liu2010oscillatory,lu2010massive,shan2010effective}.
In the latter, an effective nanosheet Hamiltonian is obtained for each number of QLs by projecting the full 3D bulk Hamiltonian onto the subspace of four $\Gamma$-point surface states that appear due to the nontrival topology and which are closest to the Fermi level.
For samples of \bismselsp with a thickness of a few QLs, these theoretical studies predict a series of topological phase transitions between the 3D and 2D models, in which the material repeatedly oscillates between a trivial phase and a QSH phase.
Focusing on the ultrathin regime between 1 and 6 QLs, as for larger thicknesses the surface-state gap becomes negligible, nanosheets of 6 QLs are predicted to show QSH behavior, while those with 5 and 4 QLs are found to be trivial, followed by another QSH phase at 3 QLs, and finally remaining in a trivial phase for 2 and 1 QLs.
However, this is at odds with state-of-the-art first-principles calculations, which predict a nonzero $\mathbb{Z}_{2}$ invariant for 3 to 5 QLs and a trivial phase for 1 and 2 QLs\ \cite{liu2010oscillatory}.
It also disagrees with a recent experiment on colloidal \bismselsp nanosheets of finite lateral dimensions, where QSH edge states were observed directly in the regime between 4 and 6 QLs\ \cite{moes2024colloidal}.
Thus, it may seem as though the applicability of the continuum model in the regime between 4 and 6 QLs can be put into question even from the point of view of qualitative predictions in \bismsel.

It thus appears that the effective four-band model obtained in this way does not provide a complete understanding of the physics in the entire regime between 1 and 6 QLs.
Here, we demonstrate that some of the oscillations predicted by the aforementioned theoretical studies are spurious, and disappear when one takes into account not only the gapped surface states closest to the Fermi level, but also the first set of bulk conduction and valence states.
The resulting eight-band Hamiltonian is not only capable of describing the physics in a broader range of momenta, but also reproduces most of the experimental findings and features of \textit{ab initio} calculations.
Our results show that the three-dimensional continuum model for topological \bismselsp can be successfully used even in the ultrathin limit of 1 QL, and also elucidates the crucial role of the bands further away from the Fermi level, which are usually neglected.

This article is organized as follows.
In {Sec.~\ref{sec:effModel}} we present a general description of \bismselsp nanosheets by starting from the 3D bulk model, and compare the features of the four-band model for the surface states with our novel eight-band model including also the bulk states closest to the Fermi level.
In {Sec.~\ref{sec:analysisTopology}} we analyze the topology of the eight-band model by studying the interplay between the different bands.
In {Sec.~\ref{sec:comparison}} we validate our model by comparing its topological properties with those calculated from first principles as well as with the most recent experimental results.
Finally, in {Sec.~\ref{sec:summConc}} we summarize our work and give an outlook on potential future research.

\section{Effective model for \bismselsp nanosheets}
\label{sec:effModel}

Our starting point is the $\vec{k} \vec{\cdot} \vec{p}$ Hamiltonian for three-dimensional \bismselsp derived in Refs.\ \cite{zhang2009topological,liu2010model}.
This model is expressed in the combined orbital-spin basis $(\ket{\mathrm{Bi}^{+}, \uparrow}, \ket{\mathrm{Se}^{-}, \uparrow}, \ket{\mathrm{Bi}^{+}, \downarrow}, \ket{\mathrm{Se}^{-}, \downarrow})$ that is closest to the Fermi surface, where $\mathrm{Bi}^{+}$ and $\mathrm{Se}^{-}$ are hybridized Bi and Se $p_{z}$ orbitals of even and odd parity, respectively.
The effective Hamiltonian reads
\begin{equation}
\label{eq:Hamiltonian3D}
    \begin{aligned}
        H(\vec{k}, k_{z}) &= \epsilon_{0}(\vec{k}, k_{z}) \mathbb{I}_{s} \otimes \mathbb{I}_{\tau} + \mathcal{M}(\vec{k}, k_{z}) \mathbb{I}_{s} \otimes \tau_{z} \\
        &\hphantom{=\,\,} + A_{1} k_{z} s_{z} \otimes \tau_{x} + A_{2} (\vec{k} \vec{\cdot} \vec{s}) \otimes \tau_{x} \, .
    \end{aligned}
\end{equation}
Here, $\epsilon_{0}(\vec{k}, k_{z}) = C + D_{1} k_{z}^{2} + D_{2} k^{2}$, $\mathcal{M}(\vec{k}, k_{z}) = M - \smash[tb]{B_{1} k_{z}^{2}} - \smash[tb]{B_{2} k^{2}}$, $\vec{s}$ and $\vec{\tau}$ are the Pauli matrices in the spin and orbital spaces, respectively, and $\mathbb{I}_{s}$ and $\mathbb{I}_{\tau}$ are the identity matrices in these respective subspaces.
Here and below, $\vec{k} \equiv (k_{x}, k_{y})$ denotes the in-plane momentum, with $k \equiv |\vec{k}|$.
We employ the parameters of the paper by Zhang \textit{et al.}\ \cite{zhang2009topological}, which have been fitted to their \textit{ab initio} calculation.
Their numerical values are ${C = \SI{-0.0068}{\electronvolt}}$, ${M = \SI{0.28}{\electronvolt}}$, ${A_{1} = \SI{0.22}{\electronvolt\nano\meter}}$, ${A_{2} = \SI{0.41}{\electronvolt\nano\meter}}$, ${B_{1} = \SI{0.10}{\electronvolt\nano\meter\squared}}$, ${B_{2} = \SI{0.566}{\electronvolt\nano\meter\squared}}$, ${D_{1} = \SI{0.013}{\electronvolt\nano\meter\squared}}$, and ${D_{2} = \SI{0.196}{\electronvolt\nano\meter\squared}}$.

We are interested in thin nanosheets of only a few QLs.
This geometry breaks the translational invariance in the $z$-direction and thus we must solve the model of Eq.\ \eqref{eq:Hamiltonian3D} after substituting $\smash[tb]{k_{z} \rightarrow {-} \iu \partial_{z}}$.
It is customary to employ hard-wall boundary conditions for the wave functions at both surfaces, $\Psi(z = \pm L_{z}/2) = 0$, with $L_{z}$ the nanosheet thickness.
To obtain a low-energy effective model, one first solves the Hamiltonian at the 2D $\Gamma$ point $k = 0$.
The solutions to this problem are described in detail in Refs.\ \cite{lu2010massive,linder2009anomalous,shan2010effective}.
One then projects the full Hamiltonian at nonzero $\vec{k}$ onto a subset spanned by these solutions.
The size of this subspace essentially determines the validity of the effective model around $k = 0$.
Only when projecting on all states, full equivalence with the higher-dimensional Hamiltonian is recovered.

Note that, of course, one can in principle also solve the full 3D model above at arbitrary $\vec{k}$, without needing to resort to an effective model (cf. FIG.~\ref{fig:comparisonFullAndEffModels} below).
However, this requires solving a boundary eigenproblem for each $\vec{k}$, which is undesirable for many purposes, e.g., computing many-body properties involving interaction matrix elements at different momenta, or computing Chern numbers or other topological invariants where the wave functions must be known over the entire Brillouin zone.
By contrast, the use of an effective model allows one to solve the boundary problem at a single momentum and thereafter simply diagonalize a (typically small) $\vec{k}$-dependent matrix.
For this reason, it is often useful to work with a reliable effective Hamiltonian whose $z$-dependence has been integrated out.

As 3D \bismselsp is a TI, the spectrum obtained from the Hamiltonian in Eq.\ \eqref{eq:Hamiltonian3D} for large thicknesses contains states that are localized at the surfaces of the nanosheet under consideration.
Their dispersions around the $\Gamma$ point form a gapless Dirac cone that crosses the semiconductor band gap.
As one decreases $L_{z}$ to a few QLs, the Dirac cone becomes gapped at $\Gamma$ as a result of the hybridization between the states localized at opposite surfaces.
We are interested in thicknesses small enough for this gap to be of observable magnitude.
It is not until a thickness of around $L_{z} \simeq \SI{6}{\nano\meter}$ that this gap becomes of the order of $\SI{1}{\milli\electronvolt}$, so we will focus on the regime $L_{z} \lesssim \SI{6}{\nano\meter}$.
Note that the thickness of a single QL is approximately $\SI{1}{\nano\meter}$\ \cite{okamoto1994bi,lind2005structure}, so that the thickness of the nanosheet in nanometers is in good approximation the number of QLs.
We also mention that, in this ultrathin limit, what is meant by ``surface states'' are those states whose dispersions evolve from the gapless Dirac point at large $L_{z}$, even though their wave functions are no longer strongly localized due to the aforementioned hybridization.
The remaining spectrum of the Hamiltonian \eqref{eq:Hamiltonian3D} at $k = 0$ consists of states whose energies are gapped for all $L_{z}$ and whose wave functions never show strong localization; we call these the bulk states.
Finally, we clarify that because our starting model is derived from a low-energy expansion, it does not contain the surface state replicas described by Kung \textit{et al.} located deep inside the bulk energy region \cite{kung2019observation}.
However, it is safe to assume that their effects can be neglected for the purposes of this article.

\subsection{Four-band model}
\label{sec:fourBandModel}

Previous works on the model considered here have always assumed that projecting on the topological surface states alone is enough to obtain an accurate low-energy model for any thickness.
However, as we show here, this introduces some issues and in particular the topology of the resulting model does not align with experimental findings and density-functional theory (DFT) calculations.

At the 2D $\Gamma$ point, the Hamiltonian decouples into two subspaces of opposite spin, and its spectrum contains four electronic surface states close to the Fermi level.
These states have well-defined spin and parity, so we denote them by $\ket{\mathrm{S}^{\pm},{\uparrow\!(\downarrow)}}$, where the label S indicates that they are surface states \footnote{The states $\ket{\mathrm{S}^{+},{\uparrow (\downarrow)}}$ and $\ket{\mathrm{S}^{-},{\uparrow (\downarrow)}}$ correspond to $\ket{\chi^{\uparrow (\downarrow)}}$ and $\ket{\varphi^{\uparrow (\downarrow)}}$ in Refs.\ \cite{lu2010massive,shan2010effective}, respectively.}.

We define the basis $(\ket{\mathrm{S}^{+},{\uparrow}}, \ket{\mathrm{S}^{-},{\downarrow}}, \ket{\mathrm{S}^{-},{\uparrow}}, \ket{\mathrm{S}^{+},{\downarrow}})$, in which the effective four-band Hamiltonian for the surface states takes the form
\begin{equation}
    \begin{aligned}
        H^{\mathrm{eff}}_{\text{4-band}, \xi}(\vec{k}) &= (E_{0} - D k^{2}) \mathbb{I}_{\sigma} + v_{\mathrm{F}} (\vec{\sigma} \times \vec{k}) \vec{\cdot} \hat{\vec{z}} \\
        &\hphantom{=\,\,} - \xi \bigg(\frac{\Delta}{2} - B k^{2}\bigg) \sigma_{z} \, .
    \end{aligned}
\end{equation}
There are two subblocks with $\xi = \pm 1$, sometimes called the hyperbola index, as the Hamiltonians $\smash[tb]{H^{\text{eff}}_{\text{4-band},\pm}(\vec{k})}$ resemble those at the $K$ and $K'$ points of graphene but shifted to the $\Gamma$ point.
Here, however, we choose to call $\xi$ the \textit{spin-orbit parity} (SOP), as the product of spin times orbital parity in each subspace is precisely $\xi$ if we identify $\uparrow (\downarrow)$ with $+1 \, (-1)$.
Furthermore, $\vec{\sigma}$ are Pauli matrices that couple the basis states $(\ket{\mathrm{S}^{+},{\uparrow}}, \ket{\mathrm{S}^{-},{\downarrow}})$ for ${\xi = +1}$ and $(\ket{\mathrm{S}^{-},{\uparrow}}, \ket{\mathrm{S}^{+},{\downarrow}})$ for $\xi = -1$, and $\mathbb{I}_{\sigma}$ stands for the identity in each of these subspaces of fixed SOP.
We note that, even though the states in the doublets $(\ket{\mathrm{S}^{+},{\uparrow}}, \ket{\mathrm{S}^{-},{\downarrow}})$ and $(\ket{\mathrm{S}^{-},{\uparrow}}, \ket{\mathrm{S}^{+},{\downarrow}})$ have opposite spin, $\vec{\sigma}$ is \emph{not} directly the physical spin operator, as the wave functions of the up and down spin states in each subspace are not equal.
Instead, they correspond to two distinct hybridized orbitals, so that $\vec{\sigma}$ is more appropriately understood as a pseudospin that mixes the orbital and spin degrees of freedom.

Each $2 \times 2$ subblock has two bands with dispersions $\smash[tb]{\varepsilon_{c,v}(k) = E_{0} - D k^{2} \pm \sqrt{(\Delta/2 - Bk^{2})^{2} + (v_{\mathrm{F}} k)^{2}}}$.
The corresponding eigenstates are
\begin{equation}
  \psi^{c,v}_{\xi \vec{k}} = \mathcal{N}^{c,v}_{\xi \vec{k}} \begin{bmatrix}
    {-}\xi \big(\frac{\Delta}{2} - B k^{2} \big) \pm \sqrt{\big(\frac{\Delta}{2} - B k^{2})^{2} + (v_{\mathrm{F}} k)^{2}} \\
    {-} \iu v_{\mathrm{F}} k_{+}
   \end{bmatrix} , 
\end{equation}
where $k_{\pm} = k_{x} \pm \iu k_{y}$ and $\smash[tb]{\mathcal{N}^{c,v}_{\xi \vec{k}}}$ is readily found by normalizing the eigenstates to unity.
The Chern number of these states is given by $\smash[tb]{\mathcal{C}^{\xi}_{c,v} = \mp \frac{\xi}{2} (\operatorname{sgn} \Delta + \operatorname{sgn} B)}$.
This is nonzero only when $\Delta$ and $B$ have the same sign.
Physically, this may be understood from the fact that the topology arises from a band inversion, which in the four-band model is driven by the combined action of $\Delta$ and $B$.
The fact that the band inversion leads to a topologically nontrivial phase is rooted in the parity flip that takes place as one goes from the $\Gamma$ point to large momenta.
Due to the inversion, the valence band around $\Gamma$ has opposite parity to that when $k \rightarrow \infty$.
This argument was first formalized by Fu and Kane\ \cite{fu2007btopological}, and in systems with inversion symmetry allows one to determine the $\mathbb{Z}_{2}$ invariant by evaluating the parity of the eigenstates at the time-reversal-invariant momenta.
This procedure will be paramount in our description of the topology in the eight-band model below.

The Hall conductivity of each subspace is then $\smash[tb]{\sigma^{xy}_{\xi}} = \smash[tb]{(e^{2}/h) \, \mathcal{C}^{\xi}_{v}}$ provided that the Fermi level stays within the gap with decreasing thickness.
The superposition of two opposite Hall conductivities causes the total $\smash[tb]{\sigma^{xy}}$ to vanish, but if the individual conductivities are nonzero the system is in a QSH phase.
However, strictly speaking this QSH effect is in terms of the pseudospin of the underlying basis, and \textit{not} in terms of the $z$-direction component of the real electronic spin.
This can be easily seen from the fact that each $2 \times 2$ subblock mixes up and down spins, contrary to the prototypical Bernevig-Hughes-Zhang model for the QSH effect, in which the two subspaces separately describe spin-up and spin-down electrons \cite{bernevig2006quantum,bernevig2006quantum2}.
As a result, there is a nontrivial spin texture along the $z$-direction given by $\langle S_{i} \rangle_{\xi \vec{k}}^{c,v}(z) = \psi^{c,v}_{\xi \vec{k}}(z)^{\dagger} S_{i} \psi^{c,v}_{\xi \vec{k}}(z)$, with $\psi^{c,v}_{\xi \vec{k}}(z) = \braket{z | \mathrm{S}^{\xi},{\uparrow}} \! \braket{\mathrm{S}^{\xi},{\uparrow} | \psi^{c,v}_{\xi \vec{k}}} + \braket{z | \mathrm{S}^{{-} \xi},{\downarrow}} \! \braket{\mathrm{S}^{{-} \xi},{\downarrow} | \psi^{c,v}_{\xi \vec{k}}}$ and $\smash[tb]{S_{i} = \frac{1}{2} s_{i} \otimes \mathbb{I}_{\tau}}$ the spin operator.
Note that $\braket{z | \mathrm{S}^{\pm},{\uparrow\!(\downarrow)}}$ are simply the wave functions corresponding to the surface states at the $\Gamma$ point introduced in the previous section, which are $z$-dependent four-component vectors in the orbital-spin basis.

\begin{figure}[!t]
    \centering
    \includegraphics[width=\linewidth]{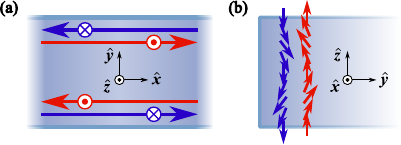}
    \caption{Sketch of the spin behavior in the QSH regime of a thin \bismselsp nanoribbon along the $x$-direction.
    \textbf{(a)} Top view of the nanoribbon.
    The long arrows along the edges show the velocity of the edge states, and the out-of-plane arrows indicate the direction of the $z$-averaged electron spin.
    Note that the average spin polarization is in general not 100\%, even though its direction of polarization is perpendicular to the nanosheet.
    \textbf{(b)} Side view of the nanoribbon.
    The small arrows show the nontrivial microscopic spin texture along the nanosheet thickness.
    This texture is such that there is a $z$-dependent tilt in the $yz$-plane.
    Its average over the nanosheet thickness vanishes in the $y$-direction, but is nonzero in the $z$-direction.
    Thus, the usual QSH picture emerges in terms of this nonvanishing vertical component, as shown in (a).
    }
    \label{fig:slab3D}
\end{figure}

Similarly, the edge states of this Hamiltonian at the boundaries of a finite sample also present a nontrivial spin texture along the vertical direction.
An analysis of these edge modes reveals that the physical spin is always perpendicular to the momentum along the edge.
More precisely, there is a nonvanishing projection in the $z$-direction, whose average over the nanosheet thickness is in general nonzero.
Furthermore, the spin in the direction perpendicular to the edge and parallel to the nanosheet has a nontrivial texture along the vertical direction, but its average over $z$ vanishes.
Finally, the spin in the direction parallel to the edge is identically zero for all $z$.
Consequently, we recover the well-known QSH picture, realized now in terms of the $z$-averaged vertical component of the real electronic spin.
We emphasize that this is not necessarily clear \textit{a priori}, given that the underlying basis mixes the up and down components as explained above.
A sketch of the situation is shown in FIG.~\ref{fig:slab3D}, where the edge is taken along the $x$-direction and thus the spin lies entirely in the $yz$-plane.
For nonzero $k_{x}$, the edge states always follow a linear dispersion, given by
\begin{equation}
\label{eq:fourBandEffEdge}
    \varepsilon^{\pm}_{\mathrm{edge}}(k_{x}) = E_{\Gamma} \pm \tilde{v}_{\mathrm{F}} k_{x} \, ,
\end{equation}
where $\tilde{v}_{\mathrm{F}} = v_{\mathrm{F}} \sqrt{\smash[b]{1 - D^{2} / B^{2}}}$ is a renormalized Fermi velocity which is lower than that of the surface states, and $E_{\Gamma}$ is the energy at the one-dimensional $\Gamma$ point $k_{x} = 0$.
We note that all of this is valid only if $|D| < |B|$, as otherwise the global energy gap disappears and no edge states are found.

The form of Eq.\ \eqref{eq:Hamiltonian3D} is indeed capable of describing the surface states of electrons in \bismselsp nanosheets, as previous experiments have successfully matched the observed band structure with that arising from the effective model\ \cite{zhang2010crossover}.
However, there is a caveat: it is crucial to realize that a faithful description of the system in terms of the four-band model requires adjusting the numerical values of the parameters in the low-energy Hamiltonian to a set of experimentally determined values.
This is not only completely \textit{ad-hoc}, but also only possible if experiments in the 2D limit are available.
Since this may not be the case in other similar materials, it is desirable to have a reliable scheme which allows 2D topological properties to be inferred solely from the 3D bulk model.
In practice, the actual theoretical prediction of the projected four-band model for \bismselsp is far from satisfactory, as the parameters obtained directly through the projection procedure give a band structure that disagrees with the experimental findings.
The most important aspect in this regard is that the topological properties appear incorrect for a relevant range of thicknesses.
The reason is that the parameters $\Delta$ and $B$, which together determine the topology, have opposite signs for 4 and 5 QLs, a regime where both experiments\ \cite{moes2024colloidal} and DFT calculations\ \cite{liu2010oscillatory} show the presence of QSH edge states.
More precisely, in this picture the model oscillates four times between a trivial and a QSH phase between 1 and 6 QLs.
Two of these oscillations are due to the gap closing, and two more seemingly take place because, while the gap remains open, the parameter $B$ changes sign and the band inversion disappears.
This phenomenon is in fact an artifact of such a continuum model: placing the model on a lattice reveals that a change in Chern number is always accompanied by a gap closing, but depending on $\Delta$ and $B$ this may happen at the edges of the Brillouin zone instead of at the $\Gamma$ point.
This is problematic, because it would indicate that the low-energy subspace we wish to study is not located at the $\Gamma$ point for a certain range of thicknesses.
As we will see, however, enhancing the model to an eight-band Hamiltonian eliminates this spurious change in Chern number.
Hence, for all thicknesses the system is still described by the physics around $\Gamma$, thus providing the physically expected picture.
Moreover, for $L_{z} \gtrsim \SI{3.23}{\nano\meter}$ it is found in the four-band model that $|D| > |B|$, which means that the valence band is not inverted and thus actually \emph{grows} in energy when $k \rightarrow \infty$.
This leads to the absence of a global gap in the spectrum, which in this system leads to no edge states even if the Chern number of the valence band is nonzero.
Once again, this is undesirable and in contradiction with experiments on \bismselsp nanosheets and DFT calculations.

\subsection{Eight-band model}

We have demonstrated that projection on the surface states alone is not enough to obtain accurate results for the low-energy physics of \bismselsp nanosheets.
We now proceed to include in this projection also the first set of bulk states that arises from solving the model at the two-dimensional $\Gamma$ point, which we denote by $\ket{\mathrm{B}^{\pm},{\uparrow\!(\downarrow)}}$.
As we explain below, this is enough to solve all issues present in the previous model.

\begin{figure}[!htb]
    \centering
    \includegraphics[width=\linewidth]{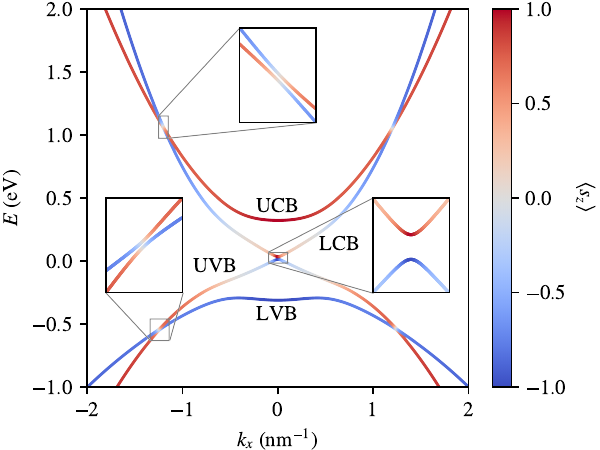}

    \caption{Band structure of a spatially extended nanosheet of 4 QLs in thickness for the subspace with SOP $\xi = +1$, as given by the eight-band model.
    The band-name abbreviations are defined in the main text.
    Here we have set $k_{y} = 0$, as the dispersion is rotationally symmetric in our $\vec{k} \vec{\cdot} \vec{p}$ model.
    The energy bands of the $\xi = {-}1$ subspace are degenerate with these ones, but their spin is opposite due to the TRS.
    Avoided crossings are found at $|k_{x}| \simeq \SI{1.2}{\nano\meter\tothe{-1}}$, at which the two valence or conduction bands exchange their spin.
    As explained in detail in the main text, this gives a nontrivial twist to the LVB and the UCB, which become topological in the effective model, while effectively undoing the twist of the UVB and the LCB visible around $|k_{x}| \simeq \SI{0.5}{\nano\meter\tothe{-1}}$ and making them trivial.
    Contrary to the four-band model for 4 QLs, the eight-band model shows a global gap, meaning that when $k \rightarrow \infty$ all conduction and valence bands go to large positive and negative energies, respectively.
    The insets show magnifications of the avoided crossings and the surface gap.}
    \label{fig:BulkBandsExtendedModel}
\end{figure}

One can show that the effective Hamiltonian always decouples into two separate subspaces, which are related by TRS.
This is a consequence of the TRS in combination with the mirror symmetry with respect to the $xy$-plane, as the latter enforces a definite parity for the individual components of the $\Gamma$-point wave functions under the operation $z \rightarrow {-}z$.
In other words, the spin-orbit parity $\xi$ is always a well-defined quantum number in the presence of both TRS and planar mirror symmetry.
For our combined eight-band Hamiltonian we choose the basis
\begin{equation}
    \begin{aligned}
        (\!&\underbrace{\ket{\mathrm{S}^{+},{\uparrow}}, \ket{\mathrm{S}^{-},{\downarrow}}, \ket{\mathrm{B}^{+},{\uparrow}}, \ket{\mathrm{B}^{-},{\downarrow}}}_{\xi = +1}, \\
        &\underbrace{\ket{\mathrm{S}^{-},{\uparrow}}, \ket{\mathrm{S}^{+},{\downarrow}}, \ket{\mathrm{B}^{-},{\uparrow}}, \ket{\mathrm{B}^{+},{\downarrow}}}_{\xi = -1}) \, .
    \end{aligned}
\end{equation}
Due to the TRS, we focus on the analysis of the first $4 \times 4$ subblock, with SOP $\xi = +1$.
Its Hamiltonian reads
\begin{equation}
    H^{\mathrm{eff}}_{\text{8-band}, \xi = +1}(\vec{k}) = \begin{bmatrix}
        H_{\mathrm{SS}} & H_{\mathrm{SB}} \\ H_{\mathrm{SB}}^{\dagger} & H_{\mathrm{BB}}
    \end{bmatrix} ,
\end{equation}
where
\begin{subequations}
    \begin{align}
        H_{\mathrm{II}} &= \epsilon_{0}^{\mathrm{I}}(\vec{k}) \mathbb{I}_{2\times2} + \begin{bmatrix}
            \mathcal{M}^{\mathrm{I}}(\vec{k}) & (A^{\mathrm{I}})^{*} k_{-} \\ A^{\mathrm{I}} k_{+} & {-}\mathcal{M}^{\mathrm{I}}(\vec{k})
        \end{bmatrix} , \\
        H_{\mathrm{SB}} &= \begin{bmatrix}
            a k^{2} & b k_{-} \\ c k_{+} & d k^{2}
        \end{bmatrix} ,
    \end{align}
\end{subequations}
with $\smash[tb]{\mathrm{I} \in \{\mathrm{S}, \mathrm{B}\}}$, $\smash[tb]{\epsilon_{0}^{\mathrm{I}}(\vec{k}) = C^{\mathrm{I}} + D^{\mathrm{I}} k^{2}}$, and $\smash[tb]{\mathcal{M}^{\mathrm{I}}(\vec{k}) = M^{\mathrm{I}} - B^{\mathrm{I}} k^{2}}$.
All parameters $\smash[tb]{(M^{\mathrm{I}}, A^{\mathrm{I}}, B^{\mathrm{I}}, C^{\mathrm{I}}, D^{\mathrm{I}}, a, b, c, d)}$ depend on the thickness $L_{z}$ and together determine the topology of the corresponding nanosheet.
It is important to realize that their numerical values are unambiguously determined from the initial set of parameters of the 3D bulk Hamiltonian, thus not requiring any further adjustments.
In FIG.~\ref{fig:BulkBandsExtendedModel} we show the band structure of the eight-band model for 4 QLs in the subspace $\xi = +1$.
There are four bands which we call the upper and lower valence or conduction band (from top to bottom: UCB, LCB, UVB, and LVB).
The topological properties can be understood by tracking the spin projection of the different bands as a function of momentum while taking care of some subtleties detailed in the next section.

\begin{figure*}[!thb]
    \centering
    \includegraphics[width=\linewidth]{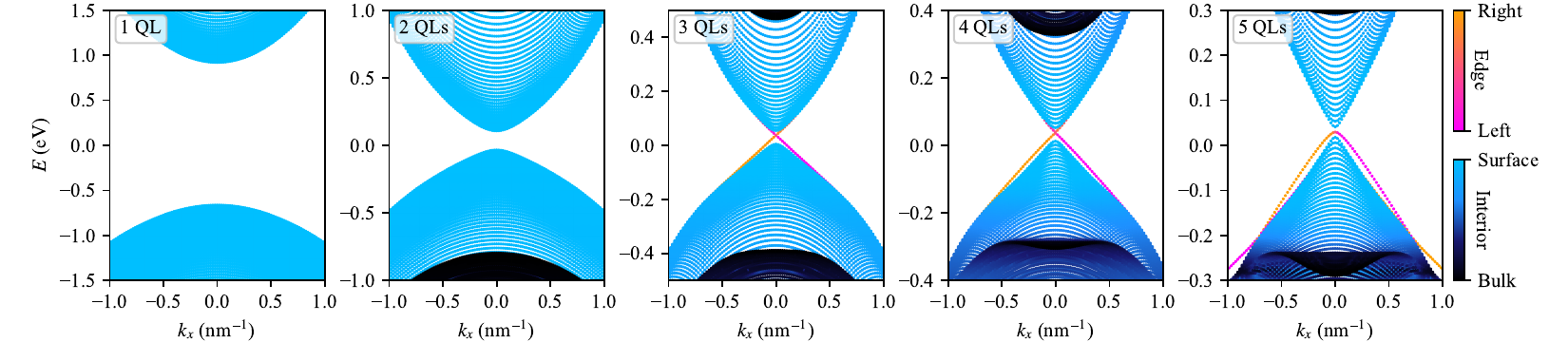}
    \includegraphics[width=\linewidth]{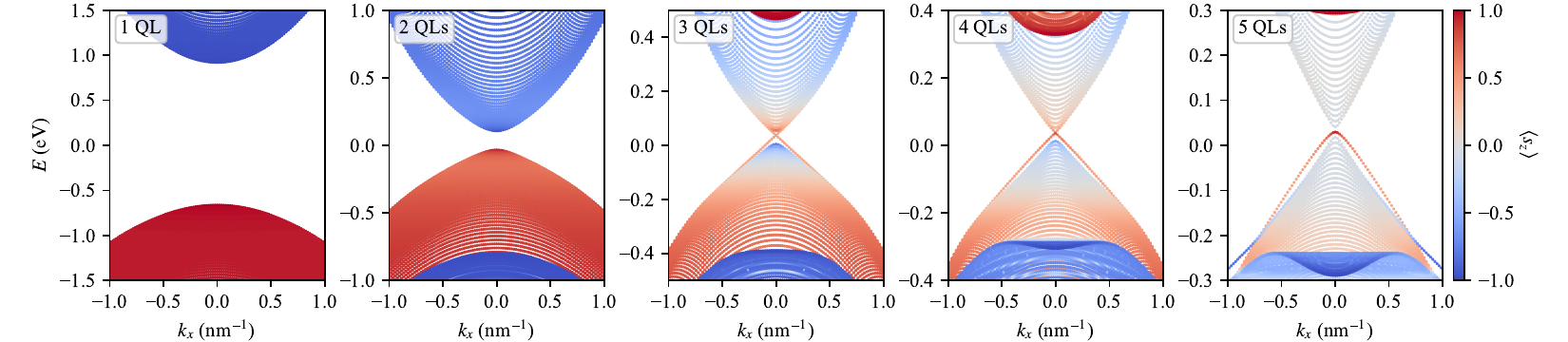}

    \caption{Band structure of a ribbon of width $L_{y} = \SI{100}{\nano\meter}$ and 1 to 5 QLs in thickness for the subspace with SOP $\xi = +1$, as given by the eight-band model.
    All parameters used arise directly from the projection of the 3D states at the $\Gamma$ point.
    The top row depicts the surface or bulk character of the bands for 1 to 5 QLs, that is, whether the corresponding states predominantly live in the outer or inner QLs, respectively.
    The edge states, only present between 3 and 5 QLs, are colored according to their localization on the ribbon.
    The bottom row depicts the spin polarization along the $z$-direction for 1 to 5 QLs.
    Despite not being fully polarized, the edge states still have a nonzero average spin in the $z$-direction, as explained in detail in Sec.\ \ref{sec:fourBandModel}.
    One must keep in mind that there is another subblock whose band structure is related by TRS to the one shown here.
    Hence, the spins and velocities of electrons in the time-reversed subspace are opposite, giving the usual QSH picture but with a partial spin polarization.}
    \label{fig:MergingEdgeBulk100nm}
\end{figure*}

We have computed the spectrum of each subspace separately on a ribbon along the $x$-direction.
The width in the $y$-direction is taken as $L_{y} = \SI{100}{\nano\meter}$ and we solve the continuum model by substituting $\smash[tb]{k_{y} \rightarrow -\iu \partial_{y}}$ and employing hard-wall boundary conditions for the wave function at the edges.
The results for 1--5 QLs are plotted in FIG.~\ref{fig:MergingEdgeBulk100nm}.
In all cases the valence bands are fully inverted, i.e., they go down in energy when $k \rightarrow \infty$, meaning that for $L_{z} \gtrsim \SI{3.23}{\nano\meter}$ the bulk states are required to obtain a fully gapped spectrum.
For 1 and 2 QLs the spectrum is devoid of edge states, as already happens in the four-band model.
However, we now find edge states not only for 3, but also for 4 and 5 QLs, in contrast to the situation in the four-band model.
In the first row of the figure we have colored the states according to their surface or bulk character, and the edge states are shown according to their localization along the width of the ribbon.
In the second row we show the states colored according to their average spin in the $z$-direction.
It is worth noting that the picture of FIG.~\ref{fig:slab3D} remains valid, except for the fact that the precise spin texture is now also influenced by the wave functions of the newly added bulk states.
We comment on the case of 5 QLs, where there seems to be no Dirac point.
This is due to the fact that the gap closing takes place at a thickness of $\SI{5.02}{\nano\meter}$, so that the last panel of FIG.~\ref{fig:MergingEdgeBulk100nm} is extremely close to the transition.
As such, the gap is so small that the upper half of the Dirac cone, which is slightly gapped due to finite-size effects, has already merged with the lowest conduction band.
Although not shown here, this can be verified by tracking the evolution of the edge conduction band at intermediate thicknesses between 4 and 5 QLs, which poses no complications in a continuous model.


\section{Analysis of the topology}
\label{sec:analysisTopology}

To determine the topological protection of the aforementioned edge states it is necessary to compute the $\mathbb{Z}_{2}$ invariant of the system, as the total $8 \times 8$ Hamiltonian lies in class AII\ \cite{chiu2016classification}.
However, the system decouples into two subspaces that get interchanged under time reversal, so in this case the $\mathbb{Z}_{2}$ classification is equivalent to a double Chern-number classification\ \cite{li2010chern}.
The topology can thus be equally determined from the Chern numbers of each subblock, as they individually break the TRS.
A distinction will have to be made between \emph{physical} Chern numbers (PCNs) and \emph{effective-model} Chern numbers (EMCNs).
The latter refer to those obtained by directly integrating the Berry curvatures of the eight-band model in the entire range of $k$, which is unbounded in our continuum model.
We will argue that they generally differ from the PCNs, i.e., the Chern numbers which should be considered physical, albeit the Hall conductivity stays the same.
These PCNs will be defined later and can be easily inferred once we have gained an intuitive understanding of the features of FIG.~\ref{fig:BulkBandsExtendedModel}.

We have numerically computed the EMCNs of the four bands of each subspace.
The total Hall conductivity is then $\smash[tb]{(e^{2} / h) \sum_{i} \mathcal{C}^{\xi}_{i}}$, where the sum runs over the occupied bands.
In the range between the gap closings at $L_{z} = \SI{2.51}{\nano\meter}$ and $L_{z} = \SI{5.02}{\nano\meter}$, we find that the EMCN of the UVB is zero, whereas that of the LVB band is nontrivial and equal to $\pm 1$ in each subblock.
The Hall conductivity is thus nonzero and we have accordingly found that edge states are present.
For $L_{z} < \SI{2.51}{\nano\meter}$ and $\SI{5.02}{\nano\meter} < L_{z} \leq \SI{6}{\nano\meter}$, we actually find that \emph{both} occupied bands of each subblock have opposite unit EMCNs.
Thus, the total Hall conductivity vanishes and there is no spin-Hall current.

\begin{figure}[!htb]
    \centering
    \includegraphics[width=\linewidth]{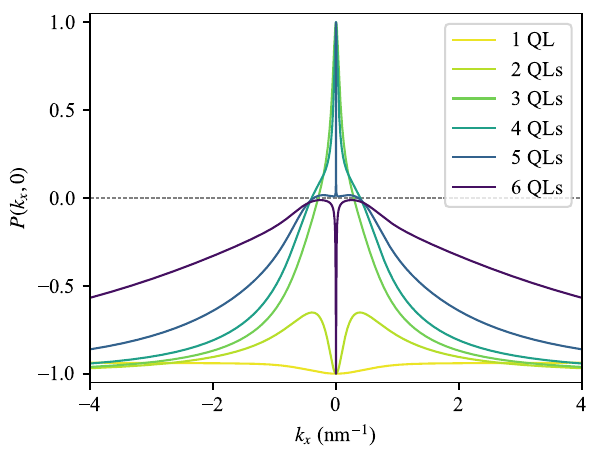}

    \caption{Pfaffian of the matrix $A(\vec{k})$ for 1 to 6 QLs in the eight-band model.
    In the cases of 3, 4, and 5 QLs, the Pfaffian changes sign twice along the contour, giving $\nu = 1$, while the other cases are trivial.
    Note that our model has rotational symmetry and thus the Pfaffian's zeros, if any, form a circle enclosing the origin of the $(k_{x}, k_{y})$-plane.
    }
    \label{fig:Pfaffian}
\end{figure}

For completeness, we have also explicitly calculated the $\mathbb{Z}_{2}$ invariant directly by analyzing the Pfaffian of the matrix
\begin{equation}
  A_{ij}(\vec{k}) = \braket{\psi_{i \vec{k}} | \Theta | \psi_{j \vec{k}}} ,
\end{equation}
which we denote by $P(\vec{k}) \equiv \operatorname{Pf}[A(\vec{k})]$.
Here, $\Theta$ is the time-reversal operator, and one takes only the eigenstates $\smash[tb]{\ket{\psi_{i \vec{k}}}}$ whose energy lies below the Fermi level.
In our eight-band model, the time-reversal operator is represented by the matrix $\Theta = {-}\iu (\sigma^{\mathrm{SOP}}_{x} \otimes \sigma^{\mathrm{SB}}_{z} \otimes \sigma_{y}) \mathcal{K}$, where $\smash[tb]{\sigma^{\mathrm{SOP}}_{x}}$ acts on the two uncoupled subspaces with opposite SOP, $\smash[tb]{\sigma^{\mathrm{SB}}_{z}}$ acts on the surface-bulk degree of freedom, $\sigma_{z}$ acts on the pseudospin defined in Sec.\ \ref{sec:fourBandModel} (generalized to surface and bulk orbitals), and $\mathcal{K}$ is the complex conjugation operator.
The topological invariant is determined by analyzing the phase of $P(\vec{k})$ along a contour $C$ that encloses precisely \emph{half} of the Brillouin zone, or in the case of our continuum model, half of the infinite momentum plane.
The $\mathbb{Z}_{2}$ invariant is then given by\ \cite{kane2005z2,bernevig2013topological}
\begin{equation}
    \nu = \frac{1}{2} \times [\text{sign changes of $P(\vec{k})$ along $C$}] \text{ mod } 2 \, .
\end{equation}
Choosing $C$ to be formed by a line integral over $k_{x}$ and an irrelevant semicircular path around the upper half plane with $\smash[tb]{k \rightarrow \infty}$, we can simply analyze $\smash[tb]{P(k_{x}, k_{y} = 0)}$.
This is plotted in FIG.~\ref{fig:Pfaffian} for 1--6 QLs, and it is found that it remains negative for 1, 2, and 6 QLs, but changes sign twice for 3 to 5 QLs.

The results for the EMCNs determined above can now be explained in an intuitive manner.
We refer to the picture of FIG.~\ref{fig:BulkBandsExtendedModel}, which shows the case of 4 QLs, and focus on the valence bands.
At the $\Gamma$ point, the gap defined by the surface states is inverted, as the spin reverses when moving away from it.
Thus, the UVB exchanges its character with the LCB.
However, at larger momenta there is now an avoided crossing (AC) between the UVB and the LVB.
This causes the spin of the UVB to flip again, while also reversing that of the LVB.
Thus, the LVB ultimately acquires a nontrivial twist when going from $k = 0$ to $k \rightarrow \infty$, while the UVB remains trivial because its twist is undone at the AC.
The same happens for 3 and 5 QLs.
As a result, there is no net twist of the UVB and its EMCN is accordingly zero, while the net spin flip of the LVB results in a nontrivial EMCN.
In contrast, for 1, 2, and 6 QLs, the closing of the gap undoes the band inversion at the origin.
However, the AC remains, resulting in the spin twisting once for every band, but oppositely between the UVB and the LVB.
The result is that the occupied bands have nonzero but opposite EMCNs, leading to a vanishing Hall conductivity, in agreement with our calculations.

In fact, this intuitive understanding becomes rigorous by virtue of the parity argument by Fu and Kane\ \cite{fu2007btopological}, because in addition to the TRS, our $\vec{k} \vec{\cdot} \vec{p}$ Hamiltonian possesses the inversion symmetry of the underlying crystal lattice.
In our continuum model there are only two time-reversal invariant momenta, namely $k = 0$ and $k = \infty$, as the momentum space now has the topology of a sphere with all points at infinity identified.
We find that the eigenstates of the effective Hamiltonian with a nonzero EMCN indeed change parity as one goes from the origin of $\vec{k}$ to infinity.
This is ultimately the same as comparing the spin direction at these points, because the product of spin and parity is well-defined in each subspace due to our definition of the SOP in Sec.~\ref{sec:fourBandModel}.

We have now understood how the direction of the spin can be used to read off the topology of the bands in the effective model.
However, one has to be careful regarding the nature of the ACs involved in the process.
While these are sources of Berry curvature and thus affect the EMCNs, they are actually an artifact of the effective model, which employs a finite number of bands.
This can be seen by including even more bands: as we increase the size of the effective model, the ACs remain very localized but shift to larger values of $k$.
In the limit of an infinite number of bands these will have disappeared and one recovers the smooth band structure arising from the solution to the 3D model at arbitrary $\vec{k}$.
However, the band inversion of the LCB and the UVB close to $\Gamma$ (at around $k_{x} \sim \SI{0.4}{\nano\meter\tothe{-1}}$ in FIG.~\ref{fig:BulkBandsExtendedModel}) remains localized in the same region no matter how many bands we include.
One must therefore conclude that this inversion is really physical.
Indeed, our $\vec{k} \vec{\cdot} \vec{p}$ model is based on the premise that the topological properties of \ce{Bi2Se3} arise from the band inversion around the $\Gamma$ point due to the spin-orbit coupling\ \cite{zhang2009topological}.
Thus, the nontrivial Berry curvature should be localized around this point only, which is precisely expressed by the spin twists in this region.

In view of the above, the effective model is physical only in the region before the ACs take place.
This means that the EMCNs do not actually correspond to the Chern numbers of the model with infinitely many bands, as they include the effect of the unphysical ACs.
To remedy this, we introduce the PCNs.
These must be calculated via the above spin-flip counting procedure, but only in the region before the first AC.
Then only the physical band inversion contributes, if present.
Note that, in any model with a finite band number, the PCNs cannot be calculated by integration of the Berry curvature as the result will not be quantized due to the truncation of the momentum space.
The PCNs, as calculated via spin flips, are guaranteed to be the same as the Chern numbers of the infinite-band model.
It then follows that for 3, 4, and 5 QLs the UVB has a unit PCN while that of the LVB is trivial, that is, the effective-model picture is essentially reversed.
Furthermore, for 1, 2, and 6 QLs all bands have trivial PCNs.
However, note that the physical Hall conductivity remains unchanged with respect to that of the effective model because it is given by a \emph{sum} of Chern numbers.
Thus, it does not matter if it is calculated with the EMCNs or the PCNs, because each AC oppositely affects both involved bands.
The $\mathbb{Z}_{2}$ invariant calculated via the Pfaffian is similarly correct in both cases, since it simultaneously includes the effect of all occupied bands.

We stress that the inclusion of the bulk states at the $\Gamma$ point is essential to reach our conclusions regarding the PCNs.
As said before, there is \emph{no} band inversion around $\Gamma$ in the effective four-band model for 4 and 5 QLs, i.e., all Chern numbers are trivial in this case.
The eight-band model is thus the minimal model that contains all the topology, which can now be understood as follows.
Firstly, there is a band inversion between the LCB and the UVB that endows these bands with nontrivial PCNs.
In the region where this inversion takes place, both of these bands are essentially fully surface-like, as seen from FIG.~\ref{fig:MergingEdgeBulk100nm}.
However, this inversion is only observed in the presence of the UCB and the LVB, which in this region are almost fully bulk-like.
We thus conclude that the nontrivial topology resides in states on the surfaces and is due to a band inversion enabled by the presence of deeper-lying states with bulk character.

\begin{figure}[!t]
    \centering
    \includegraphics[width=\linewidth]{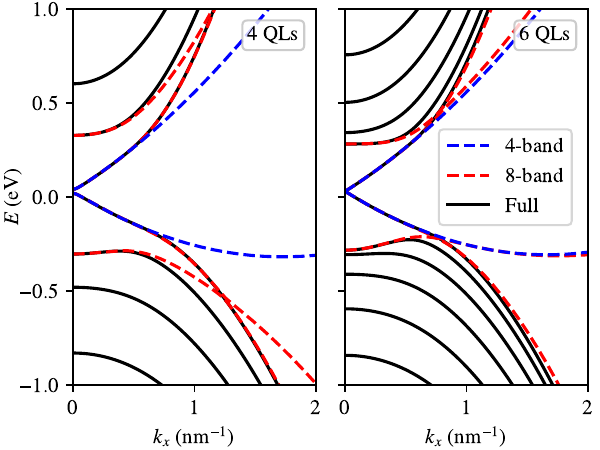}
    \caption{Comparison of the four-band and eight-band models with the exact solution of the 3D model.
    The panels show a distinctive artifact of the four-band model, namely that the valence band is not inverted for 4 QLs and higher, and thus goes up in energy for large $k$.
    In the case of 4 and 5 QLs, this is fixed by the eight-band model, whose valence bands go down in energy as desired.
    For 6 QLs and thicker, the eight-band model also presents this artifact.
    However, the topological invariant of this larger model remains the same as that of the eight-band model, which cannot be said when going from the four-band to the eight-band model.}
    \label{fig:comparisonFullAndEffModels}
\end{figure}

Finally, in TABLE~\ref{tab:topologicalProperties} we present a comparison summary between the four-band and the eight-band models.
It is apparent that the four extra bands drastically modify the properties of the model for more than 3 QLs, and must therefore be included.
We note that for 6 QLs the UVB is not inverted, leading still to the absence of a global gap, as seen in FIG.~\ref{fig:comparisonFullAndEffModels}.
We have checked that this can be fixed by including yet another set of states, but this does not change the PCNs.

\begin{table}[!t]
    \centering
        \begin{tabular}{c|c|c|c|c}
            & \multicolumn{1}{c|}{Thickness (QLs)} & \multicolumn{1}{c|}{$\hphantom{iiiii} \nu \hphantom{iiiii}$} & \multicolumn{1}{c|}{Global gap} & \multicolumn{1}{c}{Edge states} \\
            \hline
            \parbox[t]{2mm}{\multirow{6}{*}{\rotatebox[origin=c]{90}{4-band model}}}
            & 1 & 0 & \cmark & \xmark \\
            & 2 & 0 & \cmark & \xmark \\
            & 3 & 1 & \cmark & \cmark \\
            & 4 & 0 & \xmark & \xmark \\
            & 5 & 0 & \xmark & \xmark \\
            & 6 & 1 & \xmark & \xmark \\
            \hline
            \parbox[t]{2mm}{\multirow{6}{*}{\rotatebox[origin=c]{90}{8-band model}}}
            & 1 & 0 & \cmark & \xmark \\
            & 2 & 0 & \cmark & \xmark \\
            & 3 & 1 & \cmark & \cmark \\
            & 4 & 1 & \cmark & \cmark \\
            & 5 & 1 & \cmark & \cmark \\
            & 6 & 0 & \xmark & \xmark \\
        \end{tabular}
    \caption{Overview of the features of the four- and eight-band models discussed in the main text.
    Here, $\nu$ is the topological $\mathbb{Z}_{2}$ invariant, and edge states are found only when it is nontrivial \textit{and} there is a global energy gap, that is, all valence bands go to negative energies when $k$ grows large.}
    \label{tab:topologicalProperties}
\end{table}

\section{Comparison with a recent experiment}
\label{sec:comparison}

\begin{figure}
    \centering
    \includegraphics[width=\linewidth]{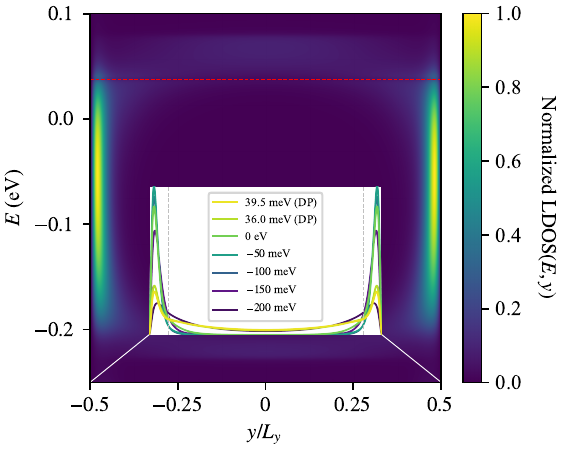}
    \caption{Local density of states of the edge modes on a ribbon of width $L_{y} = \SI{100}{\nano\meter}$.
    The states span an energy range of roughly $\SI{200}{\milli\electronvolt}$ and penetrate about $\SI{8}{\nano\meter}$ into the interior at both edges.
    The red dashed line marks the position of the Dirac point, where we find two states with a tiny gap.
    The inset shows slices of the density of states at various energies, with the ones corresponding to the Dirac point labeled as DP.
    The vertical dashed lines mark a distance of $\SI{8}{\nano\meter}$ from the edges, showing the spatial extent of the edge states that agrees with the experiment on colloidal nanosheets.}
    \label{fig:LDOS}
\end{figure}

In a recent experiment\ \cite{moes2024colloidal}, the presence or absence of QSH edge states was studied in colloidal nanocrystals of \bismselsp with thicknesses between 1 and 6 QLs.
The findings clearly show that platelets of 1 to 3 QLs do not sustain edge modes, while platelets from 4 to 6 QLs do.
A potential explanation for the 3 and 6 QL cases, which do not conform to our eight-band model, is given in the next section.
Except for this issue, our eight-band model agrees with the experimental findings, while the four-band model does not if one simply takes the parameters arising from the projection of the 3D bulk Hamiltonian.
In the experiment, the edge states are found to span a large energy window of around $\SI{500}{\milli\electronvolt}$ and penetrate around $\SI{8}{\nano\meter}$ into the interior of the crystals.
In FIG.~\ref{fig:LDOS} we have plotted the local density of edge states in the case of 4 QLs as calculated via the eight-band model.
Their penetration is about $\SI{8}{\nano\meter}$, in excellent agreement with the experimental value.
Furthermore, they span an energy range of approximately $\SI{250}{\milli\electronvolt}$.
While this is roughly only half of that observed experimentally, we note that the gap measured in the laboratory is also larger than that of our effective eight-band model \cite{zhang2010crossover}.
Hence, it is natural that the energy range spanned by the edge states is also larger than that predicted by the latter.
Nevertheless, our eight-band model still provides an insight on the underlying mechanism behind this behavior.
In FIG.~\ref{fig:MergingEdgeBulk100nm} we see that the edge states merge with the conduction band very close to the Dirac point, but that they leak remarkably deep below the top of the valence band, especially for 4 and 5 QLs.
This is attributed to a shift in the position of the Dirac point in combination with a renormalized Fermi velocity of the edge states, which around the $\Gamma$ point is slightly lower than that of the surface states.
This enables the edge modes to live well separated from the gapped surface-like bands for an energy range much larger than the zero-momentum gap.
In fact, the same conclusion can be reached via the four-band model with experimentally adjusted parameters.
For $D > 0$ (which is the case according to Ref.~\cite{zhang2010crossover}) the Dirac point shifts upwards, and the Fermi velocity in both cases becomes smaller than that of the surface states, as mentioned already after Eq.\ \eqref{eq:fourBandEffEdge}.
As a result, the edge modes of the four-band model never actually touch the valence band; the eight-band model is needed to observe the merging that takes place in FIG.~\ref{fig:MergingEdgeBulk100nm} and predict a finite energy range.

\section{Summary and outlook}
\label{sec:summConc}

In this article we have investigated the effect of the nanosheet thickness on thin films of the 3D TI \bismsel.
Whereas previous studies had already shown a dimensional transition from a 3D to a 2D TI, the theoretical description in terms of an underlying $\vec{k} \vec{\cdot} \vec{p}$ Hamiltonian was unsatisfactory.
Indeed, the emergent effective 2D model for the surface states can only describe the experimentally observed QSH phase at 4 and 5 QLs if its parameters are adjusted appropriately \textit{a posteriori}.
This amounts to disregarding the values that naturally arise from the projection in favor of a different set of parameters taken from experiment.
In this work we have shown that this becomes unnecessary upon the inclusion of bulk states further away from the Fermi level.
Our extended eight-band model precisely captures the topological properties computed via \textit{ab initio} calculations in the whole range of 1 to 6 QLs, and is close to the experimental findings except for 3 and 6 QLs.
We find that the band inversion of the surface-like bands, which determines the presence or absence of topology, can only be observed upon the inclusion of additional bulk-like bands.
This provides a new and more complete picture that is missing if one simply reparametrizes the surface bands of the four-band model.
This observation, which to the best of our knowledge has not been reported before, constitutes one of the main points of this article.

Despite the very good agreement of our eight-band model with a recent experiment and previous DFT calculations, two brief comments are in place.
Firstly, we have found that the UVB is not inverted in the regime of intermediate thicknesses of 6 QLs and beyond.
Hence, a quantitative analysis of this regime would require projection onto a higher number of states.
Note, however, that their inclusion does not further change the topology, i.e., the \textit{topological} properties can be fully explained by the eight-band model.
Secondly, our results for 3 and 6 QLs seem to clash with the recent experimental efforts involving \bismselsp nanocrystals in a way which indicates that both gap closings take place at a slightly larger thickness.
It is important to note that in this article we have employed the 3D bulk parameters as fitted from \textit{ab initio} calculations.
To further test our eight-band model it would thus be desirable to directly input the parameters from experimental studies of bulk \bismsel.
We expect that more precise knowledge of the bulk parameters will account for this discrepancy, as the values of $L_{z}$ at which the gap closes are quite sensitive to the former.

Our model is in principle not restricted to \bismsel, but also applicable to \ce{Bi2Te3} and \ce{Sb2Te3} if we use the appropriate parameter values.
Experimental research in all of these materials would thus be a useful benchmark for our eight-band model as well.
In the future we want to test our eight-band model in more general settings geared towards potential practical applications.
One interesting possibility in this regard is heterostructures, in particular those involving a superconducting substrate which induces a proximity effect on the \bismselsp slab.
Another intriguing avenue for which our combined surface-bulk Hamiltonian is especially well suited involves hybrid excitons, where an electron on the surface is coupled to a hole in the bulk, or vice-versa.
This system has been the subject of a recent experiment \cite{mori2023spin} and can be studied theoretically by combining our eight-band model with our approach to excitons in Ref.~\cite{maisel2023single}.
We hope that the considerations presented in this article will be useful for research in these and other fascinating areas of topological phases of matter.


\section*{Acknowledgments}

This work is supported by the Delta-ITP consortium and by the research program \textit{QuMat -- Materials for the Quantum Age}. These are programs of the Netherlands Organisation for Scientific Research (NWO) and the Gravitation pogram, respectively, which are funded by the Dutch Ministry of Education, Culture, and Science (OCW).

\bibliography{bib}

\end{document}